\documentstyle[preprint,aps,prc,eqsecnum]{revtex}
\begin{document}
\narrowtext
\draft
\title{Identical transitions in the strongly deformed $^{99}$Sr and
$^{100}$Sr}
\author{G. Lhersonneau,}    
\address{
Department of Physics, University of Jyv\"askyl\"a, FIN-40351 
Jyv\"askyl\"a, Finland }
\author{B. Pfeiffer, H. Gabelmann\cite{byline}, K.-L. Kratz, }
\address{
Institut f\"ur Kernchemie, Universit\"at Mainz, D-55128 Mainz, 
Germany }
\author{and the ISOLDE-Collaboration} 
\address{CERN, CH-1211 Geneva 23, Switzerland} 
\date{today}
\maketitle
\begin{abstract}
The decay of the very neutron-rich nucleus $^{100}$Rb 
has been studied by $\gamma$-spectroscopy of on-line mass-separated  
samples. Schemes for $\beta$-decay to $^{100}$Sr and $\beta$n-decay 
to $^{99}$Sr are presented. 
New sets of transitions in $^{99}$Sr and $^{100}$Sr 
with identical energies are observed. 
All identical bands so far observed in neutron-rich Sr isotopes 
obey a simple energy rule valid for even-even, odd-even and odd-odd 
bands. 
\end{abstract}
\narrowtext
\section{Introduction}
\label{sec:level1}
       
Neutron-rich isotopes with N$\geq$60 and A$\simeq$100 
are characterized by strong axial deformation. 
Strontium isotopes 
are the most deformed nuclei known so far in this region. 
Quadrupole deformations of $\beta$$\simeq$0.4 have been 
deduced for $^{98}$Sr, $^{99}$Sr and $^{100}$Sr 
from the lifetimes of the first excited states 
and from mean-square radii measured by collinear laser 
spectroscopy  \cite{OhmSr98,LSr99,LSr100,laserSreven,laserSr99}. 
According to these results, ground-state deformation 
remains constant after its sudden onset at N = 60. 
This trend might even continue at larger neutron number since 
neither the $^{101}$Sr level spacings \cite{LSr101} 
nor the 2$^{+}$-state energy of 126 keV in the N = 64 isotone   
$^{102}$Sr \cite{LSr102} indicate any large change of deformation. 
This dramatic picture has, however, been very recently contested 
on the basis of high-spin data obtained in prompt fission by the 
EUROGAM collaboration suggesting instead a more gradual increase 
of deformation with neutron number \cite{Urban100}.  
A peculiar feature of neutron-rich Sr nuclei is the occurence 
of identical bands in isotopes with $\Delta$A= 2. Transitions 
in the ground-state bands in the even-even $^{98-100}$Sr and 
in the K$^{\pi}$=3/2$^{+}$ bands in the odd-neutron  $^{99-101}$Sr 
have very close energies. A local trend of 
strong dependence of moments of inertia on deformation and the 
fact that all involved Sr isotopes have the same deformation 
seem to be the origin of these identical bands \cite{LSr100,LSr101}. 
However, the intricate mechanism is not yet understood. 

Presently, the nucleus $^{100}$Sr$_{62}$ is the most neutron-rich 
even-even Sr isotope for which experiments can yield some 
structure information. The levels in $^{100}$Sr were first observed 
at the CERN-ISOLDE facility in a $\beta$-decay study of $^{100}$Rb by 
Azuma et al. who identified the 4$^{+}\to$ 2$^{+}\to$ 0$^{+}$ 
cascade and performed the first lifetime measurement of 
the 2$^+$ state \cite{Azuma}, establishing large deformation. 
A more accurate lifetime measurement performed later by our group, 
lead to the deformation of $\beta$=0.40 \cite{LSr100}. 
In the same experiment, a 85~ns lifetime was observed for the 
1619 keV level and attributed to K-hindrance of the decay to the 
4$^+$ state by the 1202 keV transition \cite{Pfshort}.  
Recently, further members of the ground-state band up to 
I$^{\pi}=10^{+}$ were identified in prompt-fission studies 
\cite{Hamilton}. The large moment of inertia shows very little 
variation with angular momentum. Thus, $^{100}$Sr is a 
strongly deformed nucleus with properties close to the  
rotational limit. 

Yet, data on non-yrast levels in $^{100}$Sr remained scarce. 
Here we present a more comprehensive decay scheme of $^{100}$Rb 
to $^{100}$Sr, including new non-yrast levels and band structure 
built on the K-isomer.  
In addition, $\beta$-delayed neutron emission from 
$^{100}$Rb \cite{Biggy} provides a new access to levels 
in $^{99}$Sr, complementing data from 
$^{99}$Rb $\beta$-decay \cite{Pf88zz}. 

\section{Experiment and analysis}
\subsection{Experiment}
\label{sec:level2}

The $^{100}$Rb activity was produced by fission of natural uranium 
induced by 600 MeV protons, followed by on-line mass separation 
at the ISOLDE facility.  The main goal of the experiment was to 
determine the deformation of the $\beta$-decay daughter 
nucleus $^{100}$Sr from the lifetime of its 2$^+$ state at 129 keV, 
as presented in Ref.\ \cite{LSr100}. 
We now report on the result of a detailed analysis of $\gamma-$singles 
and $\gamma-\gamma-t$ coincidence measurements recorded with 
the planar Ge-detector used for the lifetime measurement 
(4.9~cm$^2$ area by 1.3~cm depth) and a coaxial Ge-detector 
of 27$\%$ relative efficiency. 
The highest energy to be recorded with the planar detector 
was set to about 1250 keV, in order to allow gating of the 
1202 keV line. 
The coaxial detector covered energies up to 4.2 MeV, which still  
is lower than the neutron separation energy S$_n$ of 
6.12 MeV for $^{100}$Sr \cite{tblis}. The low 
efficiency for coincidences where both transition energies are 
above say 1 MeV prevents the construction of a 'complete' decay scheme. 
Nevertheless, the setup allows detection of the $\gamma$-rays 
important for the discussion of the low-lying levels of $^{99}$Sr 
and $^{100}$Sr. 

\subsection{Analysis}

Since $^{100}$Rb is situated very far 
from $\beta$-stability, $\gamma$-spectra are very complex 
due to the long chain of A = 100 isobaric activities produced by 
filiation and some other A = 99 activities following 
$\beta$-delayed neutron emission from $^{100}$Rb. 

In a first step, the identification of transitions to $^{100}$Rb 
decay was made by requiring a coincidence with either the 129.2 keV 
($2^{+} \to 0^{+}$) transition in $^{100}$Sr or with the 90.8 keV 
($5/2^{+} \to 3/2^{+}$) transition in $^{99}$Sr, 
detected in the planar detector. These projections are shown 
in fig.~\ref{figspec100} and fig.~\ref{figspec99}.  
Gates were set on all the coincident transitions in the 
coaxial Ge-detector, and the new projections on the planar detector  
were analysed. Prompt coincidences with a window of about 60~ns 
and delayed coincidences were analysed to help 
placing transitions with respect to the 85~ns isomer 
in $^{100}$Sr. Many ground-state transitions have been 
placed on the sole basis of their good energy 
fit between levels established by coincidence relationships. 
The weakest of them must be regarded as tentative since there 
is a chance that such transitions also belong to 
other activities and could have been overlooked in 
previous studies. Fortunately, decay schemes 
of $^{100}$Sr and of its descendants are well known  
\cite{tblis,tristanY100,MachZr100,VogelNb100,MenzenMo100}, 
and several decays in the A = 99 mass chain were also 
re-investigated recently \cite{LZr99,LNb99,LMo99}.  

Owing to its superior energy resolution, the planar 
detector is very effective for the identification of low-energy 
transitions depopulating the K-isomer in $^{100}$Sr and of 
the band structures in $^{99}$Sr. However, its efficiency decreases 
rapidly with increasing energy. This made the placement of the  
weakest high-energy transitions uncertain, since these may be 
seen in coincidence with the 129 keV $2^{+} \to 0^{+}$ transition 
in $^{100}$Sr or the 91 or 125 keV transitions in $^{99}$Sr, 
but no longer with occasional intermediate higher-energy 
transitions. Consequently, the high-energy part of the level schemes 
remains tentative. Thus, logft values for low-lying levels 
might be strongly underestimated in these cases where feeding 
high-energy $\gamma$-rays have been overlooked.   

It has finally to be noted that the surface-ionisation ion-source 
used in our experiment produced rubidium and strontium beams 
with high efficiency. As a matter of fact, the presence of 
an intense $^{100}$Sr beam prevents the use of intensity 
balance considerations to determine the 
ground-state $\beta$-branching from $^{100}$Rb to $^{100}$Sr.  
As a consequence of the unknown amount of produced $^{100}$Rb, 
the $\beta$-delayed neutron emission probability P$_n$ 
cannot be deduced from $\gamma$-ray intensities. \\


\section{Results}
\subsection{Decay of $^{100}$Rb to $^{100}$Sr}

The levels known prior to this analysis were members 
of the ground-state band \cite{LSr100,Hamilton} and the 85~ns isomer 
at 1619 keV \cite{Pfshort}. Based on nuclear structure arguments, 
it was assumed that both the $^{100}$Rb ground state and the isomer in 
$^{100}$Sr are I$^\pi$=4$^-$ states. 
In this particular case, the delayed $\gamma$-$\gamma$ coincidences 
across the isomer allow a high sensitivity for detecting feeding 
transitions. The strong $\beta$-decay branch to the isomer is
accordingly the most reliable one deduced in this work. 
A logft of 5.6 is obtained if intensity balances are calculated 
including the tentatively placed transitions (it becomes somewhat
lower by removing them). This value is in agreement with the allowed 
character for this 4$^{-}$ to 4$^{-}$ transition.  

Transitions in the decay of $^{100}$Rb to $^{100}$Sr are listed in
table \ref{tabgam100}. The decay scheme is shown in 
fig. \ref{figsch100}. 
In the following, we present the arguments for placing the most 
important levels. 

%
%
%
The 937.8 keV level is based on the coincidence of 
the 809 keV line with the 129 keV 2$^{+}\to$ 0$^{+}$ transition.  
The intensity of this coincidence is high enough to ensure that further 
coincidences with the 288 keV line and with the most intense 
lines in the 1 MeV region have not been overlooked. 
No line of energy suitable as a ground-state transition 
is found in the singles spectra. 

The coincidence of a line in the 129 keV gate and a 
ground-state transition establish the next levels at 
1257.1 keV and 1315.4 keV with spin and parity of I$^{\pi}$=2$^+$. 
It must be mentioned here that both logft values, 5.6 and 5.8 
respectively, are clearly too low for the assumed first-forbidden 
unique transitions. Alternatives to solve the discrepancies will be
discussed in a later section. 

The levels at 1414.6, 1500.6 and 1560.6 keV are supported by 
low-energy feeding transitions from the 1619 keV (I$^\pi$=4$^-$) 
level and de-excitation to both the 129 (I$^\pi$=2$^+$) and 
417 keV (I$^\pi$=4$^+$) levels. No transitions to the ground 
state are found. Transition rates for the 
58.3, 118.0 and 204.4 keV transitions from the 85~ns isomer 
are consistent with strongly retarded dipole transitions 
or moderately delayed E2 transitions. 
Thus, I=3 of either parity or 4$^+$ are possible for these levels.  

The isomeric level at 1619 keV (t$_{1/2}$=85 ns) has been proposed 
as I$^\pi$=4$^-$ \cite{Pfshort}. The strongly hindered 1202 keV 
E1 transition to the 417 keV 4$^+$ state has to compete with 
transitions of very much lower energies to the levels at 1415, 
1501 and 1561 keV. 
Delayed coincidences indicate that the 161.8, 194.4 and 864.0 keV 
transitions are above the isomeric level. 
The prompt 162-702 keV weak coincidence  
could indicate a 702.3 keV transition above the 1781 keV level, 
but may also be due to the cross-talk of the strong 864 keV 
transition belonging to the decay of $^{100}$Y \cite{tblis}. 
A complex structure is associated with the prompt 162-194 keV 
coincidence. A contribution is due to lines placed in $^{99}$Sr. 
Another one, so far not reported, belongs to $^{99}$Zr. 
Due to these interferences a 161.8-194.4 keV 
coincidence in $^{100}$Sr remains tentative. 
We nevertheless assume the 194 and 162 keV transitions to be 
coincident and to form a band structure on top of the 1619 keV 
I$^{\pi}$=4$^{-}$ isomer for two reasons. First, the energies 
are in perfect agreement with the rotational formula 
for a 6$\to$ 5$\to$ 4 spin sequence. Second, 
$\beta$-feedings to the 1781 and 1619 keV levels are in a ratio of 
0.19(4) to 0.81(4). This is calculated neglecting conversion of 
the weak lines from the 1619 keV level and the possible feeding 
to the 1781 keV level by the 702 keV weak transition, but these
corrections remain within the quoted error. The $\beta$-feeding 
intensities are thus in excellent agreement with the theoretical 
values of 0.20 (I=5) and 0.80 (I=4) according to the Alaga 
rule \cite{Alaga}.   

The deduced $\beta$-decay branchings are inconsistent 
with the 2$^{+}$ spin and parity of the first excited 
state (logft=5.8) and of the new levels at 1257 keV (5.6) and 
1315 keV (5.8). These inconsistencies   
could be removed by lowering the ground-state spin of $^{100}$Rb 
to I=3, but contradicting the arguments for I$^{\pi}$=4$^{-}$ 
in ref.\ \cite{Pfshort} and the strong population of the 11/2$^{+}$ 
state of the $^{99}$Sr ground-state band (see next section). 
As stated above, they could be a consequence of the partial 
nature of the decay scheme. Alternatively, a yet not identified 
low-spin isomer in $^{100}$Rb could be invoked. 
This issue remains presently unsolved. Level properties are listed 
in table {\protect\ref{tablev100}}, where $\beta$-feedings have been 
calculated under the assumption of a single $\beta$-decaying parent 
nucleus. \\

\subsection{Decay of $^{100}$Rb to $^{99}$Sr}

Transitions in the decay of $^{100}$Rb to $^{99}$Sr via 
$\beta$-delayed neutron emisssion are listed in
table~\ref{tabgam99}. The decay scheme is shown in fig.~\ref{figsch99}.
There is only a partial overlap between the levels observed in 
$\beta$-decay of $^{99}$Rb \cite{Pf88zz} and in this work. 
The highest members of the K = 3/2 ground-state band of $^{99}$Sr 
are more strongly populated in $\beta$-delayed neutron decay 
of $^{100}$Rb than in $\beta$-decay of $^{99}$Rb. 

%
%
%
The new 570 keV level has transitions only to the 7/2$^+$ and 
9/2$^+$ levels of the ground-state band (g.s.b.) at 216 and 378 keV. 
The excellent energy fitting and the consistency of the 
$\vert$(g$_K$-g$_R$)/Q$_0$$\vert$ parameter extracted from 
the branching ratios suggest the 570 keV level to be 
the 11/2$^+$ member of the g.s.b., 
(see table~\ref{tabgsb1} and table~\ref{tabgsb2}). 

%
The levels at 423, 535 and 684 keV have been assigned 5/2$^+$, 
7/2$^+$ and 9/2$^+$, respectively in \cite{tblis,Pf85zz}.  
The energy of the 304.4 keV transition was unluckily stated 
as 307 keV. The energy of the last level is thus 682.1 keV. 
The highest-spin level of this band is weakly populated 
and there is no evidence for a 11/2 level. 
Odd parity logically accounts for the 
non detection of a $\gamma$-branch from the 682 keV I=9/2 level 
to the 5/2$^+$ state at 91 keV, in spite of the high efficiency 
for the expected coincidence, since the missing transition had 
to be a M2. 
The difference of populations of the g.s.b. and the excited band 
enable some insight in the decay mechanism.   
For the levels in the g.s.b. the parity change requires a p-wave 
neutron if $\beta$-decay has allowed character or a s-wave and 
first-forbidden decay. There are several ways to reach the new  
11/2$^{+}$ level at 570 keV and this is consistent with the 
experimental strong population. 
In contrast, the K=5/2 excited band could mainly be reached 
via allowed $\beta$-decay with the emission of s-wave neutrons, 
leading to a smaller range of populated spins and fewer ways 
to reach the final levels. 

Comparing the intensity of the 91-125 keV coincidences in $^{99}$Sr 
with those of various coincidence pairs in the level schemes of 
$^{99}$Y and $^{99}$Zr, one obtains a branching of 0.13(3) for 
the 125 keV line per $\beta$-delayed neutron decay of $^{100}$Rb. 
Consequently, about 50\% of the feeding bypasses the excited levels 
of $^{99}$Sr and directly feeds its 3/2$^{+}$ ground state.  
This looks to be a rather high value. It could  be 
a consequence of $\beta$-decay of $^{99}$Rb due to some, 
even modest, mass contamination or, alternatively, by additional 
population from an hypothetical low-spin isomer in $^{100}$Rb.

\section{Discussion}
\subsection{Levels in $^{100}$Sr} 

The systematics of deformed neutron-rich Sr isotopes is limited  
to $^{98}$Sr, $^{100}$Sr and $^{102}$Sr (see fig.~\ref{figsystsr}). 
%
%
This is due to the sudden onset of deformation resulting 
from the energy shifts of coexisting shapes \cite{Lshpcx}. 
Thus, the deformed 1465 keV level in $^{96}$Sr$_{58}$ corresponds to 
the $^{98}$Sr$_{60}$ ground state while the $^{96}$Sr spherical 
ground state is associated with the excited 0$^{+}$ state at 215 keV  
in $^{98}$Sr. 
We note that the lowest deformed state reported so far 
in the odd-neutron nucleus $^{97}$Sr is a K$^{\pi}$=3/2$^{+}$ 
band head at 585 keV \cite{Hamilton,LSr97}. The rate of lowering 
of the deformed minimum with N is almost constant. Extrapolating to 
$^{100}$Sr, a spherical 0$^+$ state is expected near 1.7 MeV. 
Several levels above 0.9 MeV 
with only a decay to the 2$^+$ state could be 0$^+$ states. 
However, no spin and parity assignments are possible for them. 
In contrast, the systematics of N = 62 isotones spans a larger 
number of nuclides and is much smoother, (see fig.~\ref{figsyst62}). 
Several low-spin collective levels can be followed from the 
spherical plus $\gamma$-soft $^{108}$Pd \cite{Kim} to the 
strongly deformed $^{102}$Zr \cite{Hill,Durell}, 
owing to the gradual evolution 
of their energies and decay branchings. Extrapolation to 
$^{100}$Sr predicts a rather low-lying 0$^{+}$ state for which 
the 938 keV level is a good candidate and, at somewhat 
higher energy, of two 2$^{+}$ states which could be the 
1257 and 1315 keV experimental 2$^{+}$ levels. 
These levels could be regarded as non-yrast collective deformed levels, 
i.e. as the head of the $\beta$-band, its 2$^{+}$ and the 2$^{+}$ 
$\gamma$-band head, the order of the latter two being rather 
arbitrarily chosen in fig.~\ref{figsyst62}. 
However, another interpretation was put to our 
attention by K.~Heyde \cite{Krispriv}. 
Some of these states (the 0$^+$ and a 2$^+$) could be spherical, 
corresponding to the complex structures in $^{96}$Sr and other N=58  
isotones associated with vibrational 'normal' states or 
two-particle-two-hole excitations across the Z = 40 spherical 
subshell \cite{WoodE0}. 
In this interpretation, they would be lying much lower than expected. 
This could happen if, as approaching neutron midshell, 
the deformed states nearly reach their minimum energy relative to 
the spherical ones and the change of relative energies 
becomes smaller. It might even be possible to consider  
a minimum of deformation energy before N=66. An indication 
could be the lowest excitation energies of intruder states in the 
Rh and Pd isotopes \cite{LRh111,LPd112short} which occur at N = 64. 
The difficulties to establish the nature of low-lying 0$^{+}$ states
are made apparent by the numerous studies devoted to  
$^{152}$Sm, a nucleus close to stability where detailed experiments 
are possible. In that region, a rapid shape transition also 
occurs and while the $^{152}$Sm ground-state band exhibits 
rotational structure, other interpretations have been proposed for 
the excited 0$^{+}$ band, see e.g. ref.\ \cite{Sm152a,Sm152b,Sm152c} 
and therein. 
In order to clarify the nature of the new 938, 1257 and 1315 keV
levels in $^{100}$Sr, 
it would be of great importance to search for band structure 
and to measure transition rates. 
According to systematics showing that the largest $\rho^{2}$(E0)
values have been measured in this region \cite{WoodE0}, 
an especially large value for the decay of the tentatively 0$^{+}$ 
level at 938 keV would be a signature of shape coexistence. 

%
%
The 1619 keV isomeric level has been interpreted as a 
I$^{\pi}$=4$^-$ state, based on considerations of hindrances of 
its decay to the 4$^+$ level of the ground-state band \cite{Pfshort}. 
The proposed [411]3/2$\otimes$[532]5/2 neutron configuration 
involves the lowest quasiparticles for N = 61 and N = 63 in 
this region \cite{LSr101,LZr101,LZr103,Hotchkis}.
A band built on a similar level has been identified 
in $^{102}$Zr by Durell et al. using prompt fission \cite{Durell}. 
They observed the band up to the I$^{\pi}$=9$^{-}$ level, but 
did not report a lifetime for the band head at 1821 keV. 
The postulated 2-quasiparticle configuration has been 
reproduced by Quantum Monte Carlo calculations for $^{100}$Sr 
and $^{102}$Zr performed by Capote et al.\ \cite{Capote}.  
The decrease of pairing confirms numerous calculations 
of decay properties by the QRPA model \cite{BadHonnef,whatelseQRPA}. 
The Gallagher rule \cite{Gallagher} favours coupling of antiparallel 
intrinsic spins. Thus, a 1$^-$ spin-singlet level is expected 
below the 4$^-$ spin-triplet state. 
Such a level has not been identified in this work nor in 
$\beta$-decay or prompt fission studies of $^{102}$Zr 
\cite{Hill,Durell}. It has to be mentioned that  
if the ground-state transition is the only strong decay mode,    
such a low-spin level can easily have escaped observation in 
experiments strongly relying on coincidence data. 

\subsection{Levels in $^{99}$Sr}

The new level at 570 keV is the 11/2$^+$ member of the 
previously established K = 3/2 ground-state band \cite{Pf88zz}. 
The band analysis shown in table~\ref{tabgsb2} 
yields the absolute value of the E2/M1  mixing ratio for the 
$5/2^{+}\to 3/2^{+}$ transition, $\vert \delta$(91~keV)$\vert$= 
0.171(13). 
From the lifetime of the 91 keV level, 
t$_{1/2}$= 0.58(9)~ns \cite{LSr99} we derive B(M1)= 0.040(6) W.u., 
B(E2)= 131(27) W.u. for the 91 keV transition. 
The quadrupole moment extracted from the B(E2) value is 
$\vert$Q$_{0}\vert$ = 3.27(34)~b. It corresponds to a deformation 
parameter (with second order term included) $\beta$= 0.35(4), 
a value slightly lower than in ref.~\cite{LSr99} where 
branching ratios from decay of $^{99}$Rb were used. 
This result is in modest agreement with $\beta$= 0.44(4) extracted
from a $<$r$^{2}$$>$ measurement by collinear laser spectroscopy 
\cite{laserSr99}. Part of the difference might be due to the fact that 
the $\beta$-value from the B(E2) measurement and from the laser 
spectroscopy are not exactly the same quantity. Anyway, the average 
value is close to $\beta$=0.40 which is the deformation parameter 
for $^{98}$Sr and $^{100}$Sr \cite{OhmSr98,LSr100,laserSreven}. 
From B(M1) one gets $\vert$g$_K$-g$_R$$\vert$= 0.70(6) and,  
with further assumptions of g$_s$= 0.6$\cdot$g$_{s(free)}$ 
= --2.3 and g$_R$= Z/A, one obtains the solutions 
$<$s$_z$$>$= +0.22 and --0.71. The negative value is out of range 
while the positive one is consistent with the assignment of 
the [411]3/2 orbital \cite{Pf88zz}. We note that 
$<$s$_z$$>$= 0.29 has been calculated for this orbital 
at deformation of 0.4 in $^{99}$Y where it is part of a 
three-quasiparticle isomer \cite{ram99}. 

Such an analysis cannot be performed for the proposed 
K$^{\pi}$=5/2$^{-}$ excited band due to 
the weakness of the population and the non-observation of the 
crossover transitions. 
The inertial parameter is close to 16.0 keV, which corresponds to 
96\% of the rigid-rotor estimate. This energy compression is 
systematically observed in this region for odd-parity bands, 
where it has been attributed to Coriolis mixing \cite{Hotchkis}. 
The [532]5/2 orbital is clearly favoured as being the only 
one of odd parity near the Fermi surface. The 
systematics of odd-neutron quasiparticle levels for N = 61
has to be partly revised since the 5/2$^{-}$ energy in $^{99}$Sr 
turns out to be higher than expected from a simple extrapolation 
versus deformation \cite{LZr101}. 
It would have been in principle possible to determine the parity of the 
band from branching ratios. If one assumes rather arbitrarily 
$<$s$_z$$>$= +0.5 and -0.2 for the [532]5/2 and [413]5/2 orbitals, 
respectively, the crossover transition 
from the 9/2 state is roughly ten times stronger for even parity 
than for the odd one. However, even then, its intensity 
of 0.08 relative $\gamma$-intensity units is far below 
the detection limit. 

\subsection{Identical transitions} 

Large moments of inertia and rigid rotations are exhibited by a 
number of odd or odd-odd nuclei in the A$\simeq$100 region of 
neutron-rich nuclei but $^{98}$Sr and $^{100}$Sr are the only 
even-even nuclei showing such pronounced 
features \cite{Hamilton,ram99,Hwang}.  
The $^{100}$Sr 2$^+$ state (129 keV) is among the lowest ones,  
being second only to the 2$^+$ state (126 keV) of 
$^{102}$Sr \cite{LSr102}. The moment of inertia 
extracted from the lowest members of the ground-state band 
represents 72\% of the rigid rotor value $J_{rigid}$ and the 
E(4$^+$)/E(2$^+$) ratio is 3.23.  With increasing spin, 
the g.s.b. still shows little compression with 
E(10$^+$)/E(2$^+$) = 16.3. 
These very good rotational properties are very comparable with 
those of the classical rotors in the rare earth 
and actinide regions. 

The deformed isotopes $^{98}$Sr, $^{99}$Sr and $^{100}$Sr all have 
deformation close to $\beta$= 0.4. One therefore expects transition 
energies to scale with the mass dependence of the moments of inertia, 
i.e.  $\Delta$J/J= 5/3$\cdot$$\Delta$A/A = 3.3\% for transitions in 
$^{98}$Sr and $^{100}$Sr. However, the deviation is much smaller. 
The identity of the 6$^{+}\to$ 4$^{+}\to$ 2$^{+}$ transitions was
discussed in ref.\ \cite{LSr100}. 
It has to be noted that 
the different energies of the 2$^{+}$ states, 144 and 129 keV for 
$^{98}$Sr and $^{100}$Sr, respectively, are due to shape 
coexistence in $^{98}$Sr where the low-lying 0$^+$ state at 
215 keV perturbs the ground-state band \cite{rhomix}. 
Identity of transition energies persists at higher spins \cite{Hamilton}. 
The relative deviation of the E(10$^+$)--E(2$^+$) difference of 0.23\% 
still remains much smaller than the mass scaling contribution. 
A comparable degree of identity was also reported for the few levels 
observed in the K$^\pi$=3/2$^+$ bands in the odd-neutron nuclei $^{99}$Sr 
and $^{101}$Sr \cite{LSr101}. 
 
In this work, two levels have been identified in the K$^{\pi}$= 4$^{-}$ 
two-quasiparticle band in $^{100}$Sr and the 11/2$^{+}$ level has been
added to the g.s.b. of $^{99}$Sr. This leads to observation of new 
identical transitions now in the immediate neighbours $^{99}$Sr 
and $^{100}$Sr (see fig.~\ref{figident}). 
First, the 9/2$^+$ to 5/2$^+$ energy difference of 287.2 keV in 
the [411]3/2 g.s. band of $^{99}$Sr is almost the 
same as the energies of the 4$^{+}\to$ 2$^+$ transitions 
in $^{98}$Sr and $^{100}$Sr, 289.4 and 287.8 keV 
respectively.  In this case, we note that j$_{odd-A}$ =
j$_{even-A}+1/2$ or, in other words, the transition energy remains 
the same by increasing the number of unpaired particles and the spin
of the initial level.   
It is well known that for a K= 1/2 band with decoupling parameter
$a=1$, there is a degenerate doublet structure with $\Delta$I=2 energy 
spacings like in the even--even core.  However, this simple 
explanation faces the problem that no K= 1/2 bands have been observed 
so far in the odd neighbours and there is no K= 1/2 orbital close to the
Fermi surface. 
The 11/2$^+$ to 7/2$^+$ energy difference of 353.9 keV 
in the g.s.b. of $^{99}$Sr is also fairly close to the 6$^{-}$ to 4$^{-}$ 
energy difference of 356.2 keV in the K$^{\pi}$= 4$^-$ band in $^{100}$Sr. 
This can be written as j$_{odd-A,odd-A}$ = j$_{odd-A}+1/2$. 
In both cases, the increase in transition energy due to the 
extra half-a-unit of spin (via the I(I+1) term) is counteracted 
by the increase of the moment of inertia due to the additional 
unpaired neutron. These effects turn out to compensate almost
perfectly, thus keeping the transition energies equal. A priori, 
there is no reason why this happens to be so. The increment of 
half-a-spin unit perhaps could indicate a treatment in the 
pseudospin framework as appropriate \cite{pseudo}.  
We also note that this simple rule includes the formerly reported
identical bands with $\Delta$A= 2, in which case there are no 
unpaired particles or one in both partners 
and the spins of the levels are the same. 

We note a further trend towards identical energies in Sr and Zr 
isotopes by breaking a neutron pair. 
The deformation of $^{102}$Zr is large ($\beta$=0.38 \cite{NDS102}) 
but, its moment of inertia still is significantly smaller than the one 
of $^{100}$Sr (E$_{2^+}$($^{102}$Zr)= 151.9 keV). 
Nevertheless, the energies for the transitions in the   
two-quasineutron K$^{\pi}$=4$^-$ bands, 161.8 and 194.4 keV 
for $^{100}$Sr and 159.3 and 194.4 keV for $^{102}$Zr \cite{Durell}  
are also similar, see fig.~\ref{figident}. 
Breaking a neutron pair to form the 4$^-$ isomer  
does not increase the already large moment of inertia 
of $^{100}$Sr as much as the one of $^{102}$Zr. 
Finally, the K$^{\pi}$=5/2$^-$ bands of odd-neutron nuclei 
tend to become identical at the largest deformations, as a further
increase is expected at larger N for Zr isotopes. In particular, the 
7/2$^{-}\to$ 5/2$^{-}$ energies of Zr isotopes come closer to the 
energies in Sr, see table~\ref{tabk5half}. 

%
%

\subsection{Ground states of $^{99}$Rb and $^{100}$Rb} 

According to the table of isotopes \cite{tblis}, the ground state of 
$^{99}$Rb is 5/2$^+$. This assignment is requested to account for 
sizeable $\beta$-feeding to the 7/2$^+$ state of the g.s.b. of
$^{99}$Sr. However, the $\beta$-decay branching to the 5/2$^+$ 
state is non existent. This pattern in contradiction with the Alaga
rule (the 5/2 level is calculated to be 6 times more populated
than the 7/2 state) casts some doubts about the reliability of 
intensity balances. 
In fact, systematics suggests 3/2$^{+}$. For $^{97}$Rb I= 3/2 was 
measured by laser spectroscopy \cite{Thibault} and the decay of 
$^{101}$Rb also suggested I($^{101}$Rb)=3/2 \cite{LSr101}. 
The Nilsson scheme indeed predicts the [431]3/2 proton orbital 
to be the last one occupied for Z = 37. 

According to ref.\ \cite{Pfshort} the 4$^-$ ground state of $^{100}$Rb  
originates from the coupling of this [431]3/2 proton with the 
[532]5/2 neutron which is shown experimentally to be the ground state 
of N= 63 isotones \cite{LSr101,LZr103,Hotchkis}. 
The Gallagher-Moszkowski rule \cite{GM} depresses the 
4$^-$ level with respect to its 1$^-$ partner to create the 4$^-$ 
ground state. From purely experimental 
considerations it is not possible to determine the spin of $^{100}$Rb.  
Nevertheless, odd parity is in agreement with the large 
$\beta$-feeding to the isomeric 4$^-$ level in $^{100}$Sr.  

In addition, the existence of a low-spin isomer in $^{100}$Rb 
has been conjectured as a source of feeding to low-spin 
levels in $^{100}$Sr in addition to consequences of the low
efficiency for detecting high-energy coincidence pairs.  
Another observation, less sensitive on the details of the level 
scheme, gives further support to this assumption. 
The P$_n$-value cannot be calculated since the ground-state branch 
to $^{100}$Sr cannot be measured. Nevertheless, I$^{\pi}$= 4$^{-}$ 
for $^{100}$Rb does not allow sizeable g.s. feeding. In this case, 
a P$_n$-value of 26(8)\% is calculated by comparing the intensities 
of $\gamma$-rays in $^{100}$Sr and in A = 99 activities. This is a 
subtantially higher value than obtained from direct neutron measurement 
yielding only about 6\% \cite{Biggy}. Allowing non-zero ground-state 
feeding to $^{100}$Sr, the calculated P$_n$-value is lowered. 
This way, these values could come in better agreement. 

The 1$^{-}$ level from the $\pi$[431]3/2$\otimes\nu$[532]5/2 
coupling mentioned above could indeed be associated with this 
hypothetical isomer. First-forbidden decay could populate 
2$^{+}$ states and also the tentative 0$^{+}$ state at 938 keV. 
Then, logft-values to 2$^{+}$ states would be only slightly too 
low, which is acceptable considering the uncertainties in  
$\gamma$-ray feedings by high-energy transitions. 
This is only one of the possible alternatives to generate low-spin 
and low-lying levels since there are several experimentally known 
low-lying neutron levels at N = 63. Ultimately, the existence of 
an isomer depends on the presence at low energy of I = (2,3) levels. 
The lowest lying level after the 4$^{-}$ ground state is probably   
the 3$^{+}$ state built on $\pi$[431]3/2$\otimes\nu$[411]3/2 since 
the [411]3/2 level is only a few hundreds of keV above the 
ground state, e.g. 271 keV in $^{101}$Sr \cite{LSr101} and 
259 keV in $^{103}$Zr \cite{LZr103}. The existence of a low-spin 
isomer thus critically depends on the amplitude of the 
Gallagher-Moszkowski splittings.

\section{Conclusion}

We have studied the decay of the very neutron-rich
nucleus $^{100}$Rb to its $\beta$-daughter $^{100}$Sr and 
$\beta$-n daughter $^{99}$Sr. 
The level scheme of $^{100}$Sr, 
a strongly deformed nucleus with very good rotational
properties, has been considerably extended.  Among the lowest-lying 
levels we have introduced two 2$^{+}$ states at 1257 and 1315 keV and 
a tentative 0$^{+}$ state at 938 keV, all smoothly extending the energy
systematics of N = 62 isotones. Yet their nature, members of $\beta$ and 
$\gamma$ bands or spherical intruder states, is unknown and has to be 
investigated further. 
The identification of the 5$^-$ and 6$^-$ levels of the band built on 
the 85~ns K$^{\pi}$= 4$^{-}$ isomer is suggested by transition 
energies and $\beta$-feeding intensities. 
It creates an appealing analogy with $^{102}$Zr
where a similar band is known from prompt fission. The large  
moment of inertia of the g.s. band and the low energy (1619 keV) 
of the 4$^{-}$ isomer indicate a strong reduction of pairing with 
respect to the standard estimate. These features are more pronounced 
for $^{100}$Sr than for its isotone $^{102}$Zr. However, 
the transition energies of the 4$^-$ two-quasineutron bands 
are nearly identical. 

Beta-delayed neutron decay has allowed an extension of the 
[411]3/2 ground-state band in $^{99}$Sr by adding the 
11/2$^+$ level assigned from energy fitting and branching ratios. 
Based on weaker arguments of $\beta$-n feedings 
and  $\gamma$-ray branchings, the [532]5/2 orbital has been assigned 
to the 423 keV level, head of a band whose 7/2 and 9/2 levels are 
also observed. 

The new transitions in $^{99}$Sr and $^{100}$Sr define a new 
group of transitions with very similar energies. All identical 
transitions in strontium isotopes appear to follow a simple rule 
given as j$_{n+1}$ = j$_{n}+1/2$ where $n$ is the number of 
unpaired neutrons  and $j$ the spin of the initial level. 
This extends the systematics of previously reported identical 
transitions in $\Delta$A= 2 nuclei which are a particular case 
with the same number of unpaired neutrons. 
Moreover, with increasing neutron number (i.e. presumably of 
deformation), energies of the 5/2$^{-}$ band members 
in odd-Zr isotopes tend to become close to those in Sr isotopes. 
These peculiarities are surely worth to be further investigated. \\  

\acknowledgments

This work has been supported by the German
Bundesministerium for Education and Research (BMBF), the 
German Service for Exchange with Foreign Countries \\
(DAAD), the 
Academy of Finland and the Training and Mobility of Researchers 
Program (TMR) of the European Union. 
Discussions with Dr.~K.~Heyde are gratefully acknowledged. 
It is a further pleasure to thank Drs. N. Wiehl and P.~Jones 
for recovering the data records and to Dr. J.R.~Persson for 
his help in drawing pictures. 

\newpage
\begin{figure}
\caption{
Gamma spectra recorded with the coaxial Ge detector gated by the 
129 keV ($2^{+}\to 0^{+}$) and 288 keV ($4^{+}\to 2^{+}$) 
transitions in $^{100}$Sr. 
The counts near the 288 keV peak in the 129 keV gate have been 
divided by 4. 
The scales are adjusted to yield about equal heights for the lines 
placed on top of the 4$^+$ level. Transitions included in the 
level scheme are marked by their energy if placed directly 
above the gating transition or by a closed circle if placed 
elsewhere. In the 129 keV gate a 
minor amount of contamination is due to the 130 keV transition 
($^{99}$Zr, open squares). Peaks marked with crosses are interferences 
from K X-rays of Pb due to fluorescence and lines in 
various decays without clear statistical significance mostly 
due to random coincidences. The large amount of annihilation 
radiation due to pair production reflects 
the large population of the 2$^{+}$ level by  high-energy 
transitions.  
In the 288 keV gate, a strong cross-talk due to the 469 keV 
transition in $^{99}$Nb is visible at 181 keV (diamond). 
}
\label{figspec100}       
\end{figure}

\begin{figure}
\caption{
Part of gamma spectra recorded with the coaxial detector gated by the 
91 keV ($5/2^{+}\to 3/2^{+}$) and 125 keV ($7/2^{+}\to 5/2^{+}$)  
transitions in $^{99}$Sr. 
The counts near the 125 keV peak in the 91 keV gate have been 
divided by 2. 
Transitions are marked by their energy if placed directly on the 
gating transition. Transitions belonging to the ground-state band 
are marked by closed circles and those linking the K=5/2 excited band 
to the g.s. band are indicated by open circles. 
A moderate contamination in the upper spectrum is due to a doublet
near 90 keV in $^{99}$Zr, accounting among others for a peak at 
122 keV visible on the left of the 125 keV peak. The lower spectrum 
is dominated by the coincidences due to the 125 keV line in $^{99}$Y 
which is another $7/2^{+}\to 5/2^{+}$ transitions. 
A new level in the 
ground-state band is indicated by the 192 and 354 keV lines. 
}
\label{figspec99}      
\end{figure}

\begin{figure}
\caption{
Partial decay scheme of $^{100}$Rb to $^{100}$Sr. Only the levels 
up to the K-isomer at 1619 keV and the proposed band members 
are shown since most of the high-energy transitions 
are tentatively placed.  The complete list of levels, including 
logft-values not shown here, is presented in table
{\protect\ref{tablev100}}. 
The inconsistencies of the $\beta$-feedings and proposed spins 
are discussed in the text. 
}
\label{figsch100}      
\end{figure}

\begin{figure}
\caption{
Partial decay scheme of $^{100}$Rb to $^{99}$Sr. As for 
fig.\ {\protect\ref{figsch100}} only the levels discussed 
in the text are shown. Level feedings have been calculated for 
100 $\beta$-delayed neutron decays.  
Conversion coefficients were calculated using 
$\delta$(E2/M1) mixing ratios deduced from band properties, 
see table {\protect\ref{tabgsb2}}. 
}
\label{figsch99}     
\end{figure}

\begin{figure}
\caption{
Level systematics for neutron-rich Sr isotopes  
{\protect\cite{OhmSr98,LSr100,LSr102,Hamilton,tblis,Lshpcx}} 
suggests spherical states could be expected below 2 MeV in
$^{100}$Sr.  
}
\label{figsystsr}       
\end{figure}

\begin{figure}
\caption{
Level systematics of N = 62 isotones. 
There is a gradual evolution of structure from the 
vibrator limit in the Cd-Pd region via the 
$\gamma$-soft limit dominating the structure of Ru isotopes 
to the axial symmetry limit in Zr-Sr isotopes. The nature of 
the 938, 1257 and 1315 keV levels in $^{100}$Sr 
yet remains unclear. }  
\label{figsyst62}      
\end{figure}

\begin{figure}
\caption{
Selected transitions with very similar energies 
in $^{99}$Sr, $^{100}$Sr and $^{102}$Zr. An offset has been added in
order to better show the identical transitions as a function of the 
number of quasineutrons. 
} 
\label{figident}       
\end{figure}

\newpage
\widetext
\begin{table}
\caption{
Gamma rays following the $\beta$-decay of $^{100}$Rb to $^{100}$Sr. 
Coincidences are listed for gates on the coaxial detector and 
projections onto the X-ray detector. 
Coincidences without brackets are significant at 3 standard 
deviations ($\sigma$) or better. The single occurence of a 
coincidence less significant than 2 $\sigma$ is not listed 
if not further supported by another coincidence or a 
g.s. transition. Uncertain placements are indicated by brackets. 
The intensity of unplaced transitions is calculated assuming 
they feed directly the 129 keV level.} 
\label{tabgam100}
\begin{tabular}{rrrrcl}
   E$_\gamma$    & I$_\gamma$    & Initial  & Final & \quad & Coincidences \\
  $\lbrack$keV]  & $\lbrack$\%]  & level  & level & &  \\
\tableline
  58.3 (2) & 0.20 (5) & 1619 & 1561 &   & \\ 
 106.4 (6) & 0.16 (9) & 1522 & 1415 & \tablenotemark[1] &  \\ 
 118.0 (2) & 1.1 (2)  & 1619 & 1501 &   &  \\ 
 129.2 (1) & 100      &  129 & 0    &   & 58, (118), 162\tablenotemark[2], 
        (194)\tablenotemark[2], 204, 288, 435, (1186), (1197), 1202 \\
 161.8 (2) & 4.9 (8)  & 1781 & 1619 & \tablenotemark[2] & 
        129, 288, (1202)  \\
 194.4 (3) & 0.6 (2) & (1975 & 1781) & \tablenotemark[2] & (162)  \\ 
 204.4 (3) & 1.1 (2) &  1619 & 1415  &  & (288)  \\ 
 287.8 (2) & 40.9 (38) &  417 &  129 &  & 
       (118), 129, 162\tablenotemark[2], 435, 864\tablenotemark[2], 1202 \\ 
 434.8 (2) & 3.3 (4) &  852  & 417 & \tablenotemark[3]  &  129, 288  \\
 593.8 (4) & 1.0 (4) &       &     & \tablenotemark[4]  & 129  \\
 614.8 (4) & 1.2 (4) &        &    & \tablenotemark[5]  & 129   \\   
 637.4 (3) & 1.7 (3) &  (2056 & 1419) & \tablenotemark[6] & 129  \\
 702.3 (4) & 0.8 (3) & (2483 & 1781) & \tablenotemark[7]  & 162   \\
 740.7 (5) & 0.9 (3) & (2056  & 1315) & \tablenotemark[8] & (129)  \\ 
 808.6 (3) & 3.6 (4) &  938  &  129 & \tablenotemark[9]  & 129  \\
 864.0 (3) & 2.8 (7) &  2483 & 1619 & \tablenotemark[2] & 
       129, 288, (1202) \\
 871.1 (4) & 0.5 (2) &       &      &   & (162)  \\ 
 997.5 (4) & 1.8 (4) &  1415 &  417 &   & (129), (204), (288) \\
1083.7 (3) & 2.9 (6) &  1501 &  417 &   & 129, 288  \\
1127.8 (3) & 4.0 (5) &  1257 &  129 & \tablenotemark[10] & 129  \\
1143.4 (3) & 1.7 (3) &  1561 &  417 &   &  58, 129, 288 \\
1186.2 (3) & 7.5 (8) &  1315 &  129 & \tablenotemark[10] & 129   \\
1197.4 (4) & 9.0 (15) & 1327 &  129 & \tablenotemark[9] & 129  \\ 
1201.7 (2) & 21.3 (26) & 1619  & 417 & & 
     129, 162\tablenotemark[2], (194)\tablenotemark[2], 
     288, 864\tablenotemark[2] \\
1231.0 (4) & 0.9 (5)  &  (1648 & 417) &  & (129), (288)  \\
1257.1 (3) & 9.7 (17) & 1257 &   0  & \tablenotemark[12] &  \\
1285.5 (4) & 5.4 (6) &  1415 & 129  &   & (106), 129, 204 \\
1289.5 (3) & 3.7 (5) & (1419 & 129) & \tablenotemark[13] & 129  \\
1315.3 (4) & 4.6 (8) &  1315 &   0  & \tablenotemark[11] & \\
1328.7 (4) & 1.0 (3)  & (1746 &  417) &  & 129, (288)  \\
1371.3 (4) & 8.7 (10)  & 1501 & 129 & & 118, 129  \\
1392.6 (3) & 7.6 (9) & 1522 & 129 &   & 129  \\
1431.8 (5) & 0.6 (4) & 1561 & 129 & \tablenotemark[1] & (129) \\ 
1504.0 (5) & 1.0 (5) &       &     &   & (129)  \\ 
1539.4 (7) & 1.0 (4) & 1957  &  417 &   & (129), (288)  \\ 
1699.0 (5) & 1.3 (4) & 2116  &  417 &   & (129), (288) \\
1807.8 (8) & 0.9 (5) &       &      &   & (129) \\ 
1827.8 (6) & 1.0 (5) & 1957  &  129 &   & (129)  \\ 
1883.0 (6) & 0.8 (3) &       &      &   & (129)  \\
1926.8 (3) & 8.4 (9) &  2056 &  129 &   & 129  \\
1945.9 (7) & 0.8 (4) &       &      &   & (129)  \\ 
1986.7 (4) & 1.9 (5) &  2116 &  129 &  \tablenotemark[10] & 129  \\
2055.9 (4) & 3.3 (6) &  2056 &    0 &  \tablenotemark[11] & \\
2082.2 (3) & 3.6 (7) &  2211 &  129 &  \tablenotemark[10] & 129  \\
2115.6 (3) & 3.7 (7) &  2116 &    0 &  \tablenotemark[11] & \\
2148.4 (3) & 7.4 (9) &  2278 &  129 &  \tablenotemark[10] & 129  \\
2211.6 (3) & 7.1 (12) & 2211 &    0 &  \tablenotemark[11] &  \\
2277.3 (3) & 1.8 (4) &  2278 &    0 &  \tablenotemark[11] &  \\
2336.9 (9) & 0.8 (5) &       &      &   & (129)  \\ 
2376.7 (4) & 2.1 (5) & (2506 &  129) & \tablenotemark[14] & 129  \\ 
2635.9 (8) & 0.8 (5) &       &       &   & (129) \\ 
2929.0 (9) & 0.7 (4) & (3346 & 417) & \tablenotemark[15] & (129), (288)  \\
2967.8 (7) & 1.0 (6) &  3097 & 129 &  \tablenotemark[10] & (129)  \\ 
3035.9 (8) & 1.8 (5) &  3165 & 129 &  \tablenotemark[10] & 129   \\
3097.3 (7) & 1.4 (5) &  3097 &   0 &  \tablenotemark[11] & \\
3164.9 (8) & 0.2 (1) &  3165 &   0 &  \tablenotemark[11] & \\
3187.1 (6) & 1.6 (6) & (3316 & 129) & \tablenotemark[14] &  129  \\ 
4306.4 (9) & 0.9 (5) &       &      & (129)  \\ 
4483.3 (8) & 1.2 (7) &       &      & (129) \\ 
\end{tabular}
\tablenotetext[1]
{Placement supported by weak coincidence and energy fitting.}
\tablenotetext[2]
{Transition enhanced in the delayed coincidences, due to the
   isomeric level at 1619 keV {\protect\cite{Pfshort}}.}
\tablenotetext[3]
{Transition also observed in prompt fission {\protect\cite{Hamilton}}.}
\tablenotetext[4]
{Might be due to accidental coincidence, a strong transition of 
    same energy is placed in the level scheme of $^{99}$Nb.}
\tablenotetext[5]
{Might be due to accidental coincidence, transition of same energy 
   is placed in the level schemes of $^{99}$Zr and $^{100}$Zr.}
\tablenotetext[6]
{Tentative placement, see remark for 1290 keV transition.}
\tablenotetext[7]
{Other possible placement from level 2116 keV to level 1415 keV.}
\tablenotetext[8]
{Placement by energy fitting only.}
\tablenotetext[9]
{Placement supported by a strong coincidence but no other relationship.}
\tablenotetext[10]
{Placement supported by probable ground-state transition.}
\tablenotetext[11]
{Seen in singles only while fitting as a g.s. transition.}
\tablenotetext[12]
{Only in singles while about 25\% of the intensity is due to a 
    line in $^{100}$Mo.}
\tablenotetext[13]
{Other possible placement from level 3346 keV to 2056 keV. }
\tablenotetext[14]
{Tentative due to the weak statistics and no other relationship. }
\tablenotetext[15]
{A doubly placed 1290 keV transition could further support this
    level.} 
\end{table}

\newpage
\widetext
\begin{table}
\caption{
Levels in $^{100}$Sr populated in the decay of $^{100}$Rb. The 
transition intensities used are experimental ones. The existence of an
isomer could solve unconsistencies of logft values and adopted spins, 
but it still remains speculative. It has therefore not been
attempted to decompose the $\beta$-feeding pattern into contributions
of a low and medium spin level in $^{100}$Rb.  Following this, 100 relative
$\gamma$-intensity units correspond to 57\% $\beta$-decays. 
The logft-values have been calculated with T$_{1/2}$=51~ms,
Q$_{\beta}$=13.5~MeV and P$_{n}$=6$\%$ {\protect\cite{tblis}}, 
under the assumption of no direct ground-state feeding. 
}
\label{tablev100}
\begin{tabular}{rrrl}
  Energy $\lbrack$keV] & $\beta$-feeding $\lbrack$\%$\rbrack$ & 
  log{\it ft} &  I$^{\pi}$  \\ 
\tableline
     0.0 \hspace{1cm} &    &       &  0$^{+}$  \\
   129.2 (1)  &  11.5 (24) &  5.7  &  2$^{+}$ \tablenotemark[1]  \\
   417.0 (2)  &   2.9 (26) &  6.3  &  4$^{+}$ \tablenotemark[1]  \\
   851.8 (3)  &   1.9 (2)  &  6.4  &  6$^{+}$ \tablenotemark[1]  \\
   937.8 (3)  &   2.1 (3)  &  6.4  & (0$^{+}$) \tablenotemark[2] \\
  1257.1 (3)  &   9.7 (10) &  5.6  & (2$^{+}$) \\
  1315.4 (3)  &   6.4 (6)  &  5.8  & (2$^{+}$) \\
  1326.6 (4)  &   5.1 (8)  &  5.9  &     \\
  1414.6 (3)  &   3.4 (4)  &  6.1  & (3,4)   \\
  1418.7 (3)  &   1.1 (3)  &  6.5  & \tablenotemark[3] \\
  1500.6 (3)  &   5.9 (6)  &  5.8  & (3,4) \\
  1521.7 (3)  &   4.5 (5)  &  5.9  &   \\ 
  1560.6 (3)  &   1.1 (2)  &  6.5  & (2,3,4)   \\
  1618.8 (2)  &   9.2 (15) &  5.6  & (4$^{-}$) \tablenotemark[4]  \\
  1648.0 (5)  &   0.5 (3)  &  6.9  & \tablenotemark[3] \\
  1745.7 (5)  &   0.6 (2)  &  6.8  & \tablenotemark[3] \\
  1780.6 (3)  &   2.1 (5)  &  6.2  & (5$^{-}$)  \\
  1956.8 (5)  &   1.1 (2)  &  6.5  & (2,3,4)  \\
  1975.0 (4)  &   0.4 (1)  &  6.9  & (6$^{-}$)  \\
  2056.0 (2)  &   8.2 (7)  &  5.6  & (1,2)   \tablenotemark[5]  \\
  2115.8 (2)  &   4.0 (4)  &  5.9  & 2$^{+}$      \tablenotemark[5] \\
  2211.5 (2)  &   6.1 (7)  &  5.7  & (1,2)   \tablenotemark[5] \\
  2277.4 (2)  &   5.2 (5)  &  5.7  & (1,2)   \tablenotemark[5] \\
  2482.8 (3)  &   2.0 (4)  &  6.1  &         \tablenotemark[5] \\
  2505.8 (4)  &   1.2 (2)  &  6.3  &         \tablenotemark[5] \\
  3097.2 (5)  &   1.4 (4)  &  6.2  & (1,2)   \tablenotemark[5] \\
  3165.0 (6)  &   1.2 (2)  &  6.2  & (1,2)   \tablenotemark[5] \\
  3316.4 (7)  &   0.9 (2)  &  6.3  &         \tablenotemark[5] \\
  3346.0 (9)  &   0.4 (2)  &  6.7  &         \tablenotemark[5] \\
\end{tabular}  
\tablenotetext[1] 
{Member of the ground state band {\protect\cite{Hamilton}}.} 
\tablenotetext[2]
{Spin and parity based on systematics only.}
\tablenotetext[3]
{Level based on weak evidence, not shown in Fig. {\protect\ref{figsch100}}. }
\tablenotetext[4]
{isomeric level with 85~ns {\protect\cite{Pfshort}}.}
\tablenotetext[5]
{Level not shown in Fig. {\protect\ref{figsch100}}. } 
\end{table}

\newpage
\begin{table}
\caption{
Gamma-rays in $^{99}$Sr following $\beta$-delayed neutron decay 
of $^{100}$Rb. The same conventions are used as in 
table {\protect\ref{tabgam100}}.
}
\label{tabgam99}
\begin{tabular}{rrrrcl}
   E$_\gamma$  & I$_\gamma$    & Initial  & Final & & Coincidences \\
  $\lbrack$keV] &   $\lbrack$\%]  & level & level &  & \\
\tableline
  90.8 (1) &  100     &  91  &   0 &  & 125, 162, (319) \\
 111.9 (5) & 0.6 (3)  &  535 & 423 & \tablenotemark[1] & (423)  \\  
 125.1 (2) & 40.3 (59) &  216 & 91 & \tablenotemark[2] 
        & 91, (112), 162, (192), (319)  \\
 147.6 (5) & 0.7 (4) & (682 & 535) & \tablenotemark[3] & (423)  \\
 161.9 (3) & 11.1 (13) &  378 & 216 &   & 91, (112)\tablenotemark[1], 
          125, 192, 216 \\ 
 192.0 (4) &  2.8 (12) &  570 & 378 & \tablenotemark[4] & 
          91, 125, 162 \\
 215.9 (3) & 10.8 (15) &  216 &   0 &   & 162, 319 \\
 287.2 (3) &  7.4 (13) &  378 &  91 &   & 91, (192) \\
 304.4 (4) &  0.9 (4)  &  682 & 378 & \tablenotemark[5] & 
          91, (125), (162), (287) \\ 
 318.7 (3) &  5.7 (6)  &  535 & 216 &   & 91, 125, 216 \\ 
 332.0 (2) &  7.9 (9)  &  423  & 91 &   & 91 \\
 353.9 (3) &  2.6 (5)  &  570 & 216 & \tablenotemark[4] & 91, 125, (216) \\
 422.8 (4) & 14.0 (21) &  423 &   0 &   &   \\ 
 443.8 (3) &  6.7 (8)  &  535 &  91 &   & 91 \\
 466.4 (5) &  0.3 (2)  & (682 & 216) & \tablenotemark[4] & (125)  \\ 
 646.2 (4) &  0.9 (4)  & (862 & 216) & \tablenotemark[4] & 91, 125, (216)  \\ 
 683.7 (4) &  3.9 (13) & 1106 & 423 &   &  91, (125), (332), (422) \\ 
 764.0 (3) &  5.7 (8)  &  855 &  91 &  \tablenotemark[6] & 91 \\
 777.4 (3) &  1.8 (4)  &  994 & 216 &   & 91, (125) \\
 846.8 (3) &  2.1 (5)  & 1063 & 216 & \tablenotemark[4] & 91, 125, (216)  \\
 854.7 (4) &  6.1 (8)  &  855 &   0 &  \tablenotemark[7] &  \\ 
 902.9 (3) &  4.9 (6)  &  994 &  91 &  \tablenotemark[6] & 91 \\
 936.0 (3) &  1.9 (4)  & 1152 & 216 &   & 91, 125, (216) \\
 965.1 (5) &  1.8 (4)  & 1182 & 216 & \tablenotemark[4] & (91), (125)  \\
 971.4 (9) &  1.0 (5)  & (1063 & 91) & \tablenotemark[4] & (91) \\
 981.3 (4) &  2.3 (5)  & 1072 &  91 &  \tablenotemark[6] & 91 \\
 993.7 (5) &  2.0 (4)  &  994 &   0 &  \tablenotemark[7] &  \\
1015.0 (4) & 2.1 (5)  & 1106 & 91 &            & 91 \\
1060.9 (7) & 1.0 (5)  & 1152 & 91 & \tablenotemark[8]  & 91  \\
1072.1 (4) & 2.8 (7)  & 1072 &  0 &  \tablenotemark[7] &  \\
1090.1 (5) & 1.0 (5)  & 1182 & 91 & \tablenotemark[4] & (91)  \\
1104.9 (4) & 3.1 (7)  & 1196 & 91 & \tablenotemark[9] & 91   \\
1112.0 (7) & 0.8 (4)  & (1328 & 216) & \tablenotemark[4] & (91), (125)  \\ 
1149.8 (8) & 1.0 (5)  & (1241 &  91) & \tablenotemark[4] & (91)  \\
1211.1 (9) & 0.5 (3)  & (1427 & 216) & \tablenotemark[4] & (91), (125)  \\
1291.8 (9) & 0.7 (4)  & (1383 &  91) & \tablenotemark[4] & (91)  \\
\end{tabular}
\tablenotetext[1] 
{The transition can also be placed from the 682 keV to 570 keV
  levels.} 
\tablenotetext[2] 
{About half of the observed intensity comes from the decay of 
 $^{99}$Sr which causes a large error on the calculated value. } 
\tablenotetext[3] 
{Transition placed by energy fitting.}
\tablenotetext[4] 
{Transition not reported in $\beta$-decay of 
 $^{99}$Rb {\protect\cite{Pf85zz}}. }
\tablenotetext[5] 
{Transition reported as 307.0 keV due to a misprint in 
  ref.\ {\protect\cite{Pf85zz}}. }
\tablenotetext[6] 
{Placement supported by a probable g.s. transition. }
\tablenotetext[7] 
{Seen in singles only while fitting as a g.s transition. }
\tablenotetext[8] 
{Transition masked in singles by the 1059.5 keV line 
  in $^{100}$Zr. }
\tablenotetext[9] 
{The existence of a g.s. transition is 
  unclear due to the line at 1197.4 keV in $^{100}$Sr. }
\end{table}

\newpage
\begin{table}
\caption{
Level energies of the K = 3/2 ground-state band of $^{99}$Sr. 
The symbols refer to fits with the leading term in I(I+1)-K$^2$ 
only $(a)$, with terms in first and second order in I(I+1)-K$^2$ 
$(b)$ and with leading term plus signature term $(c)$. 
}
\label{tabgsb1}
\begin{tabular}{rrrrr}
   & \multicolumn{4}{c}{Level energies [keV]} \\ 
  I   &  Exp. &   a   &     b   &   c  \\
\tableline
 5/2  &   90.8 &   89.4 &   90.8 &   90.3 \\ 
 7/2  &  215.9 &  214.5 &  216.8 &  214.9 \\ 
 9/2  &  377.9 &  375.5 &  377.0 &  378.1 \\
11/2  &  569.8 &  572.1 &  570.1 &  570.5 \\
\end{tabular}
\end{table}

\begin{table}
\caption{
Analysis of the K = 3/2 ground-state band of $^{99}$Sr according to 
the formalism shown in ref. {\protect\cite{ram99}}.  
$(a)$ the average value of $\vert$(g$_K$-g$_R$)/Q$_0$$\vert$ = 
0.215(15) barn$^{-1}$ is used to deduce $\vert$$\delta$(91)$\vert$= 
0.171(13). 
}
\label{tabgsb2}
\begin{tabular}{rccc}
  $\gamma$-ray & I$_{i}\to$ I$_{f}$   
  &  $\vert$$\delta$(I $\to$ I-1)$\vert$ 
  &  $\vert$(g$_K$-g$_R$)/Q$_0$$\vert$ [$b^{-1}$]   \\
\tableline     
  90.8 &   $5/2^{+}\to 3/2^{+}$  &  0.171(13) &   $(a)$  \\  
 125.1 &   $7/2^{+}\to 5/2^{+}$  &  0.164(16) &  0.212(21) \\ 
 161.9 &   $9/2^{+}\to 7/2^{+}$  &  0.160(18) &  0.215(24) \\
 192.0 &  $11/2^{+}\to 9/2^{+}$  &  0.131(38) &  0.254(74) \\
\end{tabular}
\end{table}

\begin{table}
\caption{
Energies of transitions in the K = 5/2 bands in N = 61 and 63 
Sr and Zr nuclei. 
}
\label{tabk5half}
\begin{tabular}{lccl}
  Nucleus &  E($7/2^{-}\to 5/2^{-}$) &  E($9/2^{-}\to 7/2^{-}$) & \\ 
          &  $\lbrack$keV]           &  $\lbrack$keV]           & \\
\tableline     
 $^{99}$Sr  & 111.9 & 147.6 & {\small this work} \\ 
 $^{101}$Sr & (111.6) &     & {\protect\cite{LSr101}} \\
 $^{101}$Zr & 104.4 & 146.6 & {\protect\cite{LZr101,Hotchkis}} \\  
 $^{103}$Zr & 109.4 & 146.6 & {\protect\cite{Hotchkis}}  \\ 
\end{tabular}
\end{table}
 
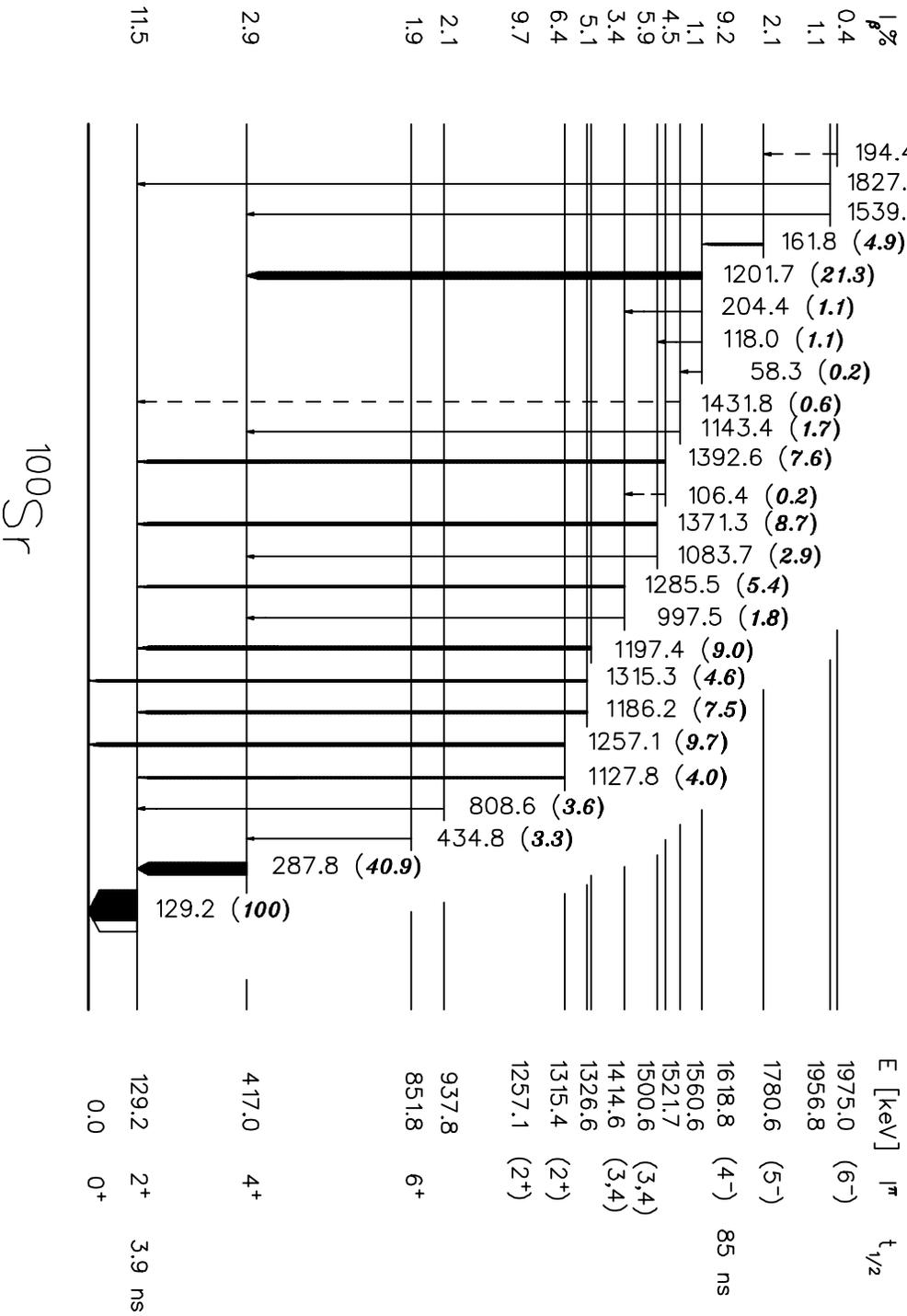
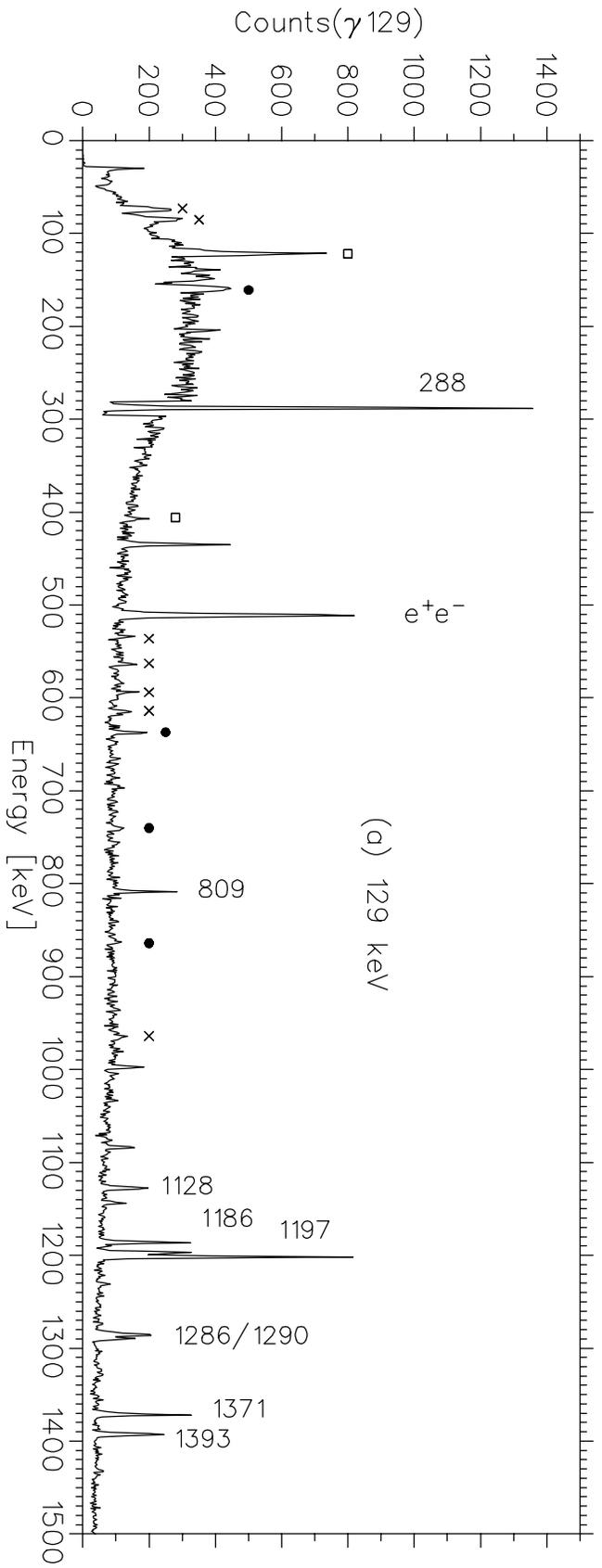
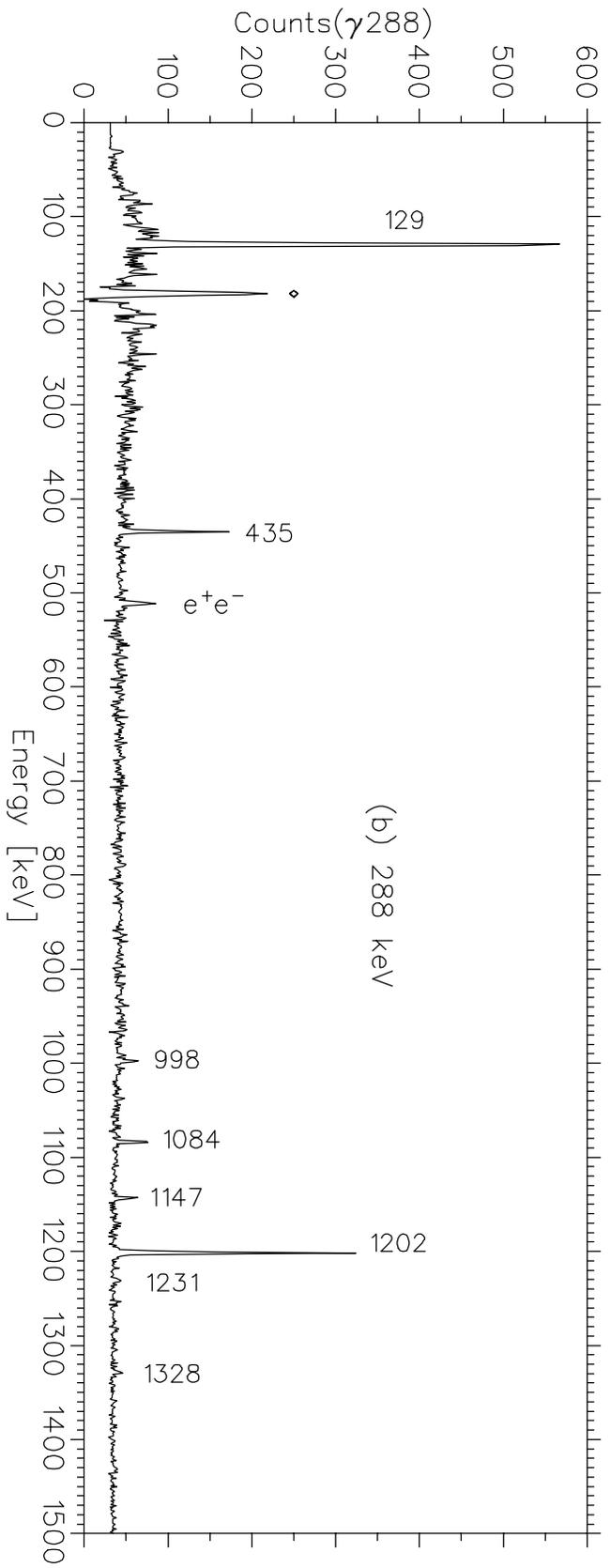
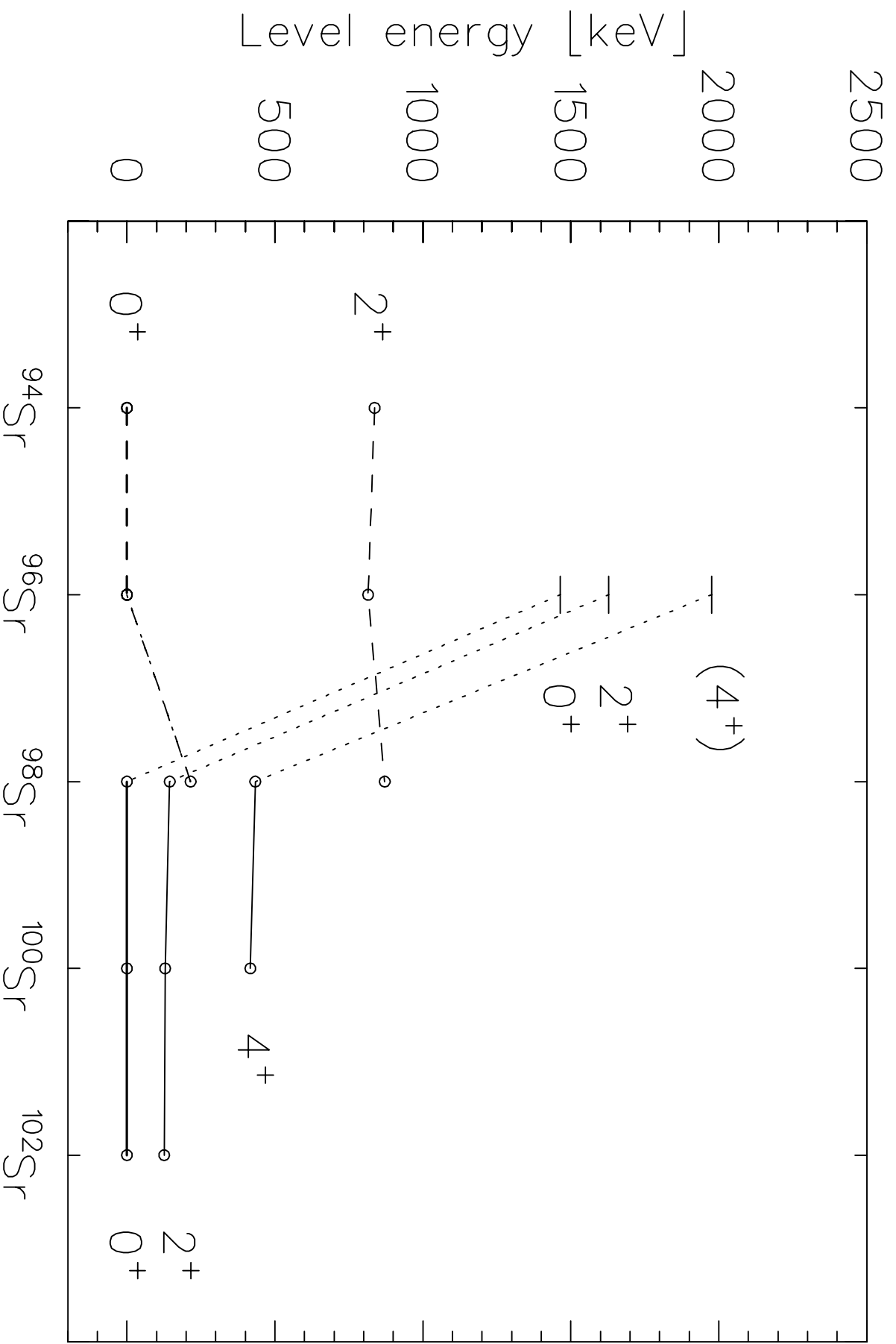
\end{document}